\documentclass[aps,superscriptaddress,twocolumn,twoside,floatfix,prl,nofootinbib,a4paper]{revtex4-2}

\pdfoutput=1

\usepackage{times}
\usepackage{amsfonts}
\usepackage{amsmath}
\usepackage{amssymb}
\usepackage{amsthm}
\usepackage{multirow}
\usepackage[normalem]{ulem}
\usepackage[T1]{fontenc}
\usepackage{float}
\usepackage{mathtools}
\usepackage{bm}
\usepackage[UKenglish,cleanlook]{isodate}   
\usepackage{mathrsfs}

\usepackage{physics}

\usepackage{xcolor}
	\definecolor{carmine}{RGB}{150,0,24}

\usepackage{natbib}
\usepackage[colorlinks=true,linkcolor=blue,citecolor=magenta,urlcolor=blue]{hyperref}

\usepackage{caption}
\usepackage{graphicx}
\usepackage{tikz-cd}

\begin{document}


\title{Adaptive advantage in entanglement-assisted communications}

\author{Jef Pauwels}
\author{Stefano Pironio}
\affiliation{Laboratoire d'Information Quantique, Universit\'e libre de Bruxelles (ULB), Belgium}
\author{Emmanuel Zambrini Cruzeiro}
\affiliation{Instituto de Telecomunica\c{c}\~{o}es, Instituto Superior T\'ecnico, Lisboa Portugal.}
\author{Armin Tavakoli}
\affiliation{Institute for Quantum Optics and Quantum Information -- IQOQI Vienna, Austrian Academy of Sciences, Austria}
\affiliation{Atominstitut,  Technische  Universit{\"a}t  Wien, Vienna,  Austria}

\date{10th of March, 2022}

\begin{abstract}
Entanglement is known to boost the efficiency of classical communication. In distributed computation, for instance, exploiting entanglement can reduce the number of communicated bits or increase the probability to obtain a correct answer. Entanglement-assisted classical communication protocols usually consist of two successive rounds: first a Bell test round, in which the parties measure their local shares of the entangled state, and then a communication round, where they exchange classical messages. 
Here, we go beyond this standard approach and investigate adaptive uses of entanglement: we allow the receiver to wait for the arrival of the sender's message before measuring his share of the entangled state. We first show that such adaptive protocols improve the success probability in Random Access Codes. Second, we show that once adaptive measurements are used, an entanglement-assisted bit becomes a strictly stronger resource than a qubit in prepare-and-measure scenarios.
We briefly discuss extension of these ideas to scenarios involving quantum communication and identify resource inequalities.
\end{abstract}

\maketitle

\begin{figure}[t!]
\includegraphics[width=\columnwidth]{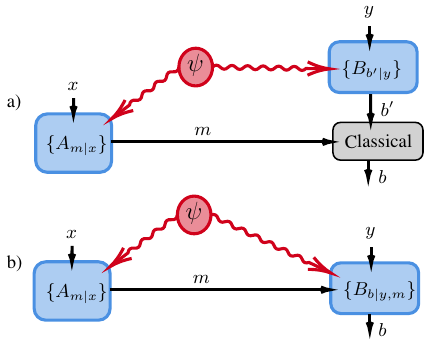}
	\caption{(a) Non-adaptive scenario for classical communication. Alice and Bob use their inputs $(x,y)$ to perform local measurements on a shared state $\psi$. Alice's outcome, $m$, is relayed to Bob who uses it to map his measurement output $b'$ into the final output $b$. (b) Adaptive scenario for classical communication. Alice uses $x$ to measure her share and relays the output $m$ to Bob who uses it to select the quantum measurement which produces the final output $b$.}\label{Scenario}
\end{figure}

\textit{Introduction.} Entanglement alone can neither be used to transfer information nor increase the capacity of a classical channel. Nevertheless, it underlies quantum advantages in classical communications. Using shared entanglement, distant parties may compute a distributed function while exchanging fewer bits of classical communication (see e.g.~\cite{Cleve1997, Buhrman1998, Raz1999, Brassard2003, Yossef2004, Laplante2018}) or with higher success probability (see e.g.~\cite{Brukner2002,  Brukner2003, Pawlowski2010, Muhammad2014, Tavakoli2016, Ho2021, Vaisakh2021, Pauwels2021, Tavakoli2021}). Underlying these classical communication advantages is the ability of entanglement to produce quantum nonlocality \cite{Buhrman2010, Brukner2004, Buhrman2016, Zukowski2017, Tavakoli2020}. Typically, the parties first measure their local share of the entangled state, generating correlations that violate a certain Bell inequality, and then use the nonlocal correlations as quantum advice on how to encode and decode the classical messages (see the review~\cite{Review}). The success of such protocols can frequently be reduced to the degree of Bell violation (as in e.g.~\cite{Buhrman2001, Brukner2004, Zukowski2017}). Such conventional protocols, illustrated in Figure~\ref{Scenario}a,  consist in viewing the Bell test and the classical communication as separate parts of the protocol.

However, nonlocality can be used to enhance communication in a more sophisticated way. In general, it can  be used adaptively, without separating the protocol into a Bell test step and a classical communication step. The receiver may delay measuring his share of the entangled state until a message arrives from another party. This message can then be used as an advise when choosing the measurement settings, as illustrated in Figure~\ref{Scenario}b. This is no longer a standard Bell test, because the sender's information is actively exploited by the receiver.

In this work, we investigate the relationship between such adaptive schemes and standard non-adaptive ones. We focus on the simplest and ubiquitous communication scenario of Figure~\ref{Scenario}, featuring a single sender, Alice, and a single receiver, Bob. 

\textit{Adaptive and non-adaptive entanglement-assisted correlations.} We describe a general sender-receiver protocol for entanglement-assisted (EA) classical communication. Alice and Bob share the entangled state $\psi$. Alice selects an input $x\in[X] \equiv \{1,\dots,X\}$ and associates a quantum measurement $\{A_{m|x}\}_m$, where the output $m$ serves as the message which has an alphabet size $D$. Bob uses $m$ together with his  input $y\in[Y]$ to select a quantum measurement $\{ B_{b|y,m}\}_b$ with outcome $b \in [B]$. The resulting probability distribution becomes 
\begin{equation}
 p(b|x,y) = \sum_{m=1}^{D} \bra{\psi} A_{m|x} \otimes B_{b|y,m} \ket{\psi}  \,.\label{eq:eacc}
\end{equation}
Interestingly, this may be interpreted as a coarse-graining of Bell scenario correlations $p(m,b|x,(y,z))$ where Alice inputs $x$ and outputs $m$ while Bob inputs both $y$ and $z\in[D]$ and outputs $b$ \cite{Anubhav2021}. We recover \eqref{eq:eacc} by post-processing the probabilities in the Bell scenario into $p(b|x,y)=\sum_{m=1}^{D}\sum_{z=1}^Dp(m,b|x,(y,m))$. In this sense, all EA advantages are powered by nonlocality. Consequently, if $\psi$ has a Bell local model, it can never enable an EA advantage (see also \cite{Vieira2022}).

Non-adaptive protocols are more restricted. While Alice, as above, uses $x$ to select a measurement $\{A_{m|x}\}_m$ with outcome $m$, Bob uses $y$ (but not $m$) to select a measurement  $\{ B_{b'|y}\}_b'$ with outcome $b'$. 
Once the message $m$ is received, Bob applies a post-processing function to map $(m,b')$ into his final outcome $b$ (see Figure~\ref{Scenario}a). We simplify Bob's description by absorbing the post-processing into his measurement: view the quantum device as outputting a list of outcomes $\vec{b}\equiv (b_1,\dots,b_D)$, where $b_k \in [B]$, for every given $y$ and the entry $b_k$ represents the final output when the message is $m=k$. This is possible because all entries in $\vec{b}$ can be known simultaneously, as they are classically computed from the outputs of the Bell test. Hence we can formally write in \eqref{eq:eacc}, $B_{b|y,m}=\sum_{\vec{b}} \delta_{b_m,b} B_{\vec{b}|y}$, which leads to the correlations, 
 \begin{align} 
	p(b|x,y) &=  \sum_{m=1}^D \sum_{\vec{b}} \delta_{b_m,b}\bra{\psi} A_{m|x} \otimes B_{\vec{b}|y} \ket{\psi} \,. \label{eq:bellcc} 
\end{align}

Using the connection to nonlocality,  we can bound the set of distributions $p(b|x,y)$ under adaptive and non-adaptive strategies using the NPA hierarchy \cite{Navascues2007, Navascues2008} (as already noted in \cite{Tavakoli2021} for the general, adaptive, case). Note that the evaluation of the semidefinite relaxations for \eqref{eq:bellcc} is more demanding than for \eqref{eq:eacc} due to Bob's large number of outputs. However, this can be remedied \cite{Note} by exploiting symmetrisation methods \cite{Gatermann2004, Ioannou2021, Tavakoli2019}. 

The constraint $B_{b|y,m} = \sum_{\vec{b}} B_{\vec{b}|y} \delta_{b_m,b}$ relating \eqref{eq:eacc} to \eqref{eq:bellcc} is equivalent to the commutation relations 
\begin{equation} 
	\forall y,b,b',m,m': ~~~[B_{b|y,m},B_{b'|y,m'}] = 0. \label{eq:commutators} 
\end{equation}
The necessity of commutation is immediate and the sufficiency follows from commuting measurements, for each $y$, admitting the mother POVM $B_{\vec{b}|y}=B_{b_1|y,1}\ldots B_{b_D|y,D}$. This provides a simple way the check whether a protocol is adaptive. It also allows to bound the set of non-adaptive $p(b|x,y)$ by adding the commutation constraints \eqref{eq:commutators} in the NPA hierarchy for \eqref{eq:eacc}.

We now proceed to show that adaptive protocols enable communication advantages over their non-adaptive counterparts for the natural Random Access Code (RAC) task  \cite{Ambainis1999, Ambainis2008, tritRAC}, which is a primitive for many quantum information purposes (see e.g.~\cite{Pawlowski2009, Aguilar2018, Tavakoli2018, Carmeli2020}). However, let us first mention the simpler scenarios where Bob only has one input ($Y=1$): it is known that entanglement still provides advantages over classical models \cite{Frenkel_2022}. One may notice  that no such advantage is possible in non-adaptive protocols because these only involve a single quantum measurement for Bob and classical post-processing, but nonlocality, which is necessary for an advantage, is impossible to generate with only one setting \cite{Massar2003}.

\begin{figure}[t!]

	\includegraphics[width=\columnwidth]{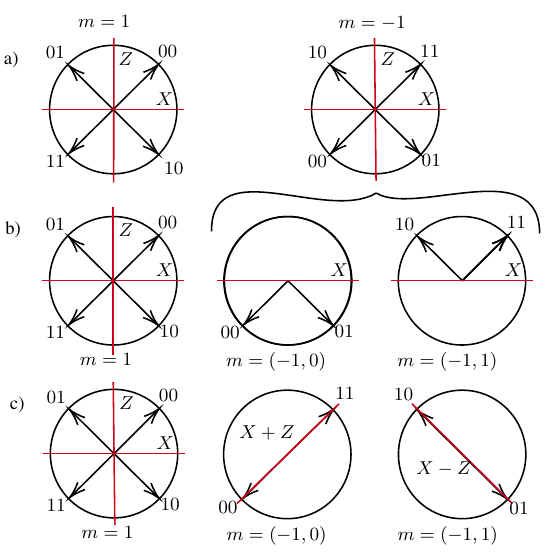}
	\caption{RAC strategies based on EA bits or trits when the shared state is $\ket{\phi^+}$. Alice measures the observable $\frac{(-1)^{x_1} Z +(-1)^{x_2} X}{\sqrt{2}}$, obtains $a=\pm 1$ and sends a message $m$ to Bob. The black arrows illustrate the Bloch vectors of Bob's local qubit after Alice's measurement for each value of $(x_1,x_2)$ and $m$. Red lines represent Bob's measurements. (a) Optimal (non-adaptive) EA bit stategy, where $m=a$. (b) A natural non-adaptive EA trit strategy, where $m=+1$ if $a=+1$ and $m=(-1,x_1)$ if $a=-1$. (c) Optimal (adaptive) EA trit strategy, where $m=+1$ if $a=+1$ and $m=(-1,x_1\oplus x_2)$ if $a=-1$.}\label{FigStrategy}
	\end{figure}

\textit{Adaptive advantage in Random Access Codes.} In a RAC task, Alice holds two bits $x \equiv (x_1,x_2) \in \{00,01,10,11\}$ that she compresses into a message of dimension $D$ that is sent to Bob. Bob gets an input $y\in [2]$ and he must produce an output $b\in\{0,1\}$ constituting a guess for the value of $x_y$. Assuming the inputs $x$ and $y$ are chosen uniformly, the average probability of a success is
\begin{equation}\label{rac}
	p_{\rm win} = \frac{1}{8} \sum_{x_1,x_2,y}p\qty(b=x_y|(x_1,x_2),y) \, .
\end{equation}

A bit message ($D=2$) achieves at most $p_\text{win}^\text{bit}=\frac{3}{4}$. This value is increased to $p^\text{EA bit}_\text{win}=\frac{1}{2}\left(1+\frac{1}{\sqrt{2}}\right)\approx 0.8535$ using pre-shared entanglement through the following CHSH-based protocol \cite{Buhrman2001}. Let Alice and Bob share the maximally entangled state $\ket{\phi^+}=\frac{1}{\sqrt{2}} ( \ket{00} + \ket{11} )$. Alice measures her local qubit with the observable $\frac{(-1)^{x_1} Z +(-1)^{x_2} X}{\sqrt{2}}$ and gets an outcome $m=\pm1 $ that she relays to Bob. This effectively prepares a local qubit state for Bob, represented in Figure~\ref{FigStrategy}a by the black arrows for the different possible values of $(x_1,x_2)$ and $m$. Bob measures his local qubit in the $Z$ ($X$) basis when $y=1$ ($y=2$) and outputs $b=1-\delta_{m,b'}$, where $b'\in\{\pm 1\}$ is the outcome of his local measurement. It is straightforward to verify that this procedure yields the claimed success probability. Note that this protocol is non-adaptive, as Bob's choice of measurement does not depend on Alice's message $m$. Since $p^\text{EA bit}_\text{win}\leq\frac{1}{2}\left(1+\frac{1}{\sqrt{2}}\right)$  for any EA bit protocol \cite{Tavakoli2021}, adaptive protocols offer no advantage in this case.

Therefore, we now consider trit communication ($D=3$), for which $p^\text{trit}_\text{win}=\frac{7}{8}$ \cite{Ambainis2008}. Now, we can readily improve the EA bit protocol by having Alice communicate not only the outcome of her local measurement, but also some information about her input: if Alice gets the output $+1$, she sends $m=+1$, if she gets the output $-1$, she sends $m=(-1,x_1)$, see Figure~\ref{FigStrategy}b. Bob behaves as before, except that if he has the input $y=1$ and he receives $m=(-1,x_1)$, he  discards the information obtained from his measurement and instead directly outputs the correct value $x_1$. It is straightforward to verify that this modified procedure yields an improved success probability of $p_\text{win}=(5+3\sqrt{2}/2)/8\approx 0.8902$.

We can further improve this protocol by having Alice perform measurements of the form $(-1)^{x_1} \cos\theta Z +(-1)^{x_2} \sin\theta X$ for some angle $\theta$. As information about $x_1$ is directly provided by Alice, we choose $\theta>\pi/4$ to increase Bob's chances of correctly guessing $x_2$ when measuring in the $X$ direction. Choosing $\cos\theta=1/\sqrt{5}$  is optimal, yielding  $p_\text{win}=\frac{5+\sqrt{5}}{8}\approx 0.9045$.

The protocol just described is non-adaptive and (at worst) nearly optimal, as we find using the NPA hierarchy at level 2,
\begin{equation}\label{tritbound}
	p^\text{NA EA trit}_\text{win} \leq 0.9082 \,,
\end{equation}
for the most general non-adaptive protocol

Now, we show that there exists an adaptive protocol, based on an EA trit, that outperforms this bound. Again starting from the EA bit protocol, we modify Alice's message as follows: if she gets output $+1$, she sends $m=+1$, if she gets output $-1$, she sends $m=(-1,x_1\oplus x_2)$. The corresponding reduced states of Bob, depending on $(x_1,x_2)$ and $m$, are depicted in Figure~\ref{FigStrategy}c. As before, when Bob receives message $m=+1$, he performs the measurements $Z$ or $X$, depending on his input. However, when the message is $m=(-1,c)$, he can perfectly guess Alice's input by performing the measurement $Z+(-1)^cX$. This strategy is adaptive, since Bob must wait for Alice's message before deciding which measurement to perform. The success probability is easily computed to be
\begin{equation}
	p^\text{A EA trit}_\text{win} = \frac{1}{4}\left(3+\frac{1}{\sqrt{2}}\right)\approx 0.9268 \label{eq:raceacc} \, ,
\end{equation}
This considerably exceeds the bound \eqref{tritbound} for non-adaptive protocols. We verified that \eqref{eq:raceacc} is actually optimal through the NPA hierarchy \cite{Tavakoli2021}.

We consider another setting where adaptive measurements yield an advantage. 
 
\textit{Simulation of qubit correlations through EA bits.}
Entanglement is often used to trade quantum communication for classical communication. The paradigmatic teleportation protocol shows that one qubit can be simulated by two EA bits. However, if one is merely interested in simulating the outcomes of projective measurements on an arbitrary communicated qubit, one EA bit is sufficient \cite{hyperbits}.

We first show that this result can be extended to general, possibly non-projective, measurements for Bob. Contrary to \cite{hyperbits}, however, our construction requires adaptive measurements. We then show that this feature is necessary: non-adaptive EA bit protocols cannot simulate non-projective correlations.

To simulate qubit correlations, it is  sufficient to focus on extremal states and measurements. The former are pure states, represented by unit Bloch vectors  $\vec{n}_x$. The latter are sets of rank-1 operators, which can be written $M_{b|y}=\frac{1}{2}\left(|\vec{v}_{by}|\openone+\vec{v	}_{by}\cdot\vec{\sigma}\right)$, where $\sum_b |\vec{v}_{by}|=2$ (normalisation) and $\sum_{b} \vec{v}_{by}=0$ (completeness). From Born's rule, we have $p(b|x,y)=\frac{1}{2}\left(|\vec{v}_{by}|+\vec{n}_x\cdot \vec{v}_{by}\right)$. To simulate $p(b|x,y)$ using an EA bit, let Alice and Bob share the state $\ket{\phi^+}$ and let Alice measure her qubit in the direction $\vec{n}'_x=(n_X,-n_Y,n_Z)$ on the Bloch sphere. On Bob's side, this prepares the qubit state $\pm \vec{n}_x$ depending on Alice's result $m=\pm 1 $. Upon receiving the message $m$, Bob performs the measurement $M_{y,m}$ defined by the operators $M_{b|y,m}=\frac{1}{2}\left(|\vec{v}_{by}|\openone+m \vec{v}_{by}\cdot\vec{\sigma}\right)$, which recovers $p(b|x,y)$.

Note that when Bob's measurements are projective, i.e., the outcomes are binary and $|\vec{v}_{by}|=1$, the measurements $M_{y,+1}$ and $M_{y,-1}$ represent the same measurement up to flipping the outcome which can be implemented non-adaptively. However, when Bob's outcomes are non-binary and the measurements non-projective, $M_{y,+1}$ and $M_{y,-1}$ are generally not jointly measurable and the protocol is adaptive.

We now prove that adaptive protocols are necessary to simulate arbitrary qubit prepare-and-measure correlations. For this, consider the following correlation function
\begin{multline}\label{facet}
	\mathcal{F}=-p(1|11) + p(1|21) + p(1|31) - p(1|12) - p(1|22) \\ + p(1|32)- p(2|12) + p(2|22) - p(2|32)\, ,
\end{multline}
corresponding to a correlation scenario $p(b|x,y)$ where Alice has three preparations ($x\in[3]$) and Bob two measurements $(y\in[2]$) with, respectively, two and three outcomes. The inequality $\mathcal{F}\leq 2$ is a facet of the polytope of correlations $p(b|x,y)$ realisable with bit messages and shared randomness.

For any qubit strategy where Bob performs projective measurements, we have $\mathcal{F}\leq \sqrt{5}$. To show this, first notice that by linearity of $\mathcal{F}$, we may without loss of generality focus on pure states, $\abs{\vec{n}_x}=1$, and rank-1 projective measurements $\frac{1}{2}\left(\openone\pm \vec{v}_{y}\cdot\vec{\sigma}\right)$, in which case Bob's three-valued measurement for $y=2$ is necessarily of one of the three types $\{M_{1|2},M_{2|2},0\}$, $\{M_{1|2},0,M_{2|2}\}$ or $\{0,M_{1|2},M_{2|2}\}$. We consider the first case, the other two can be treated analogously. The correlation function $\mathcal{F}$ then reduces to 
\begin{align}
	\mathcal{F}&=-p(1|11) + p(1|21) + p(1|31) - 2p(1|22) + 2p(1|32) -1\nonumber\\
	&=-\frac{1}{2}+\frac{1}{2}\left[-\vec{n}_1\cdot\vec{v}_1+\vec{n}_2\cdot\left(\vec{v}_1-2\vec{v}_2\right)+\vec{n}_3\cdot\left(\vec{v}_1+2\vec{v}_2\right)\right]\nonumber\\
	&\leq \frac{\left[||\vec{v}_1-2\vec{v}_2|| + ||\vec{v}_1+2\vec{v}_2||\right]}{2}\leq \sqrt{||\vec{v}_1||^2+||2\vec{v}_2||^2}\leq \sqrt{5}\,.\nonumber
\end{align}

For general measurements, however, we can achieve $\mathcal{F}= \frac{9}{4}$ using the following strategy: Alice's Bloch vectors are $(1,0,0)$, $(\frac{-1}{2},0,\frac{\sqrt{3}}{2})$, and $(\frac{-1}{2},0,\frac{-\sqrt{3}}{2})$, and Bob performs the (non-projective) measurements corresponding to $\vec{v_{11}}=-\vec{v_{21}}=(-1,0,0)$, $\vec{v}_{12}=(\frac{7\cos\theta_+}{8},0,\frac{7\sin\theta_+}{8})$, $\vec{v}_{22}=(\frac{7\cos\theta_-}{8},0,\frac{7\sin\theta_-}{8})$, $\vec{v}_{32} = (\frac{1}{4},0,0)$ with $\theta_{\pm} = -\pi \pm \arctan(4\sqrt{3})$. This both exceeds the projective bound and can be shown to be optimal \cite{Navascues2015}.

For protocols based on an EA bit, we used the NPA hierarchy to obtain an upper-bound on $\mathcal{F}$, in both the non-adaptive and adaptive cases. In both cases, we were able to find explicit strategies that match the upper-bounds. We find the tights bounds $\mathcal{F}\leq \sqrt{5}$ and $\mathcal{F}\leq \frac{9}{4}$, respectively, matching the bounds for qubit correlations for projective and general measurements. This shows that non-adaptive EA bit protocols cannot simulate a general qubit protocol. 

Combined with the result that some non-adaptive EA bits cannot be simulated by qubit protocols \cite{Pawlowski2010, Tavakoli2021}, we conclude that an EA bit is a strictly stronger resource. We also conclude that there is no strict ordering between a non-adaptive EA bit and an unassisted qubit. Note that when Bob only has one input, all singlet-assisted bit correlations can be classically simulated with two bits \cite{Frenkel_2022}.

\textit{Scenarios with quantum communication.}  The option of Bob to adapt to Alice's message naturally also extends to scenarios where the message itself is a quantum state. For general adaptive EA quantum communication \cite{Tavakoli2021, Pauwels2021}, the correlations $p(b|x,y)$ take the form
\begin{equation}
	p(b|x,y) = \Tr \qty( \qty(\$^{A\rightarrow M}_x \otimes \openone^B)[\psi]\, B^{MB}_{b|y})\, , \label{eq:eaqc}
\end{equation}
where $\psi$ is Alice and Bob's shared state, $\$^{A\rightarrow M}_x$ is a quantum channel that Alice applies on her share of $\psi$  which outputs a quantum system $M$ of dimension $D$ which is then sent to Bob, and $B^{MB}_{b|y}$ is the joint measurement of Bob on his local share of $\psi$ and on Alice's incoming message. 

In contrast, in a non-adaptive strategy, Bob would measure the message $M$ and his local system $B$ independently:
\begin{equation}
	B^{MB}_{b|y} = \sum_{b_1,b_2} p(b|b_1,b_2) B^{M}_{b_1|y}\otimes B^{B}_{b_2|y}\,.
\end{equation}

The quantum situation is richer than the classical one and in between the above two extremes, there are many intermediate situations that are also adaptive, even though they do not involve joint measurements. For instance, Bob could first measure his share $B$ of the state and use the outcome to advise his choice of measurement on $M$, i.e.~$B^{MB}_{b|y} = \sum_{b'}  B^{M}_{b|b'}\otimes B^{B}_{b'|y}$, or conversely first measure $M$ to advice the measurement on $B$, i.e.~$B^{MB}_{b|y} = \sum_{b'}  B^{M}_{b'|y}\otimes B^{B}_{b|b'}$. 

Whereas for classical communication an adaptive advantage was not immediately obvious, in the quantum case it is straightforward that general adaptive strategies lead to considerable advantages. The paradigmatic example is dense coding \cite{Bennett1992}, where Alice can communicate two $D$its to Bob, encoded in an EA qu$D$it. This means that in the task of minimal error state discrimination, where Alice has $X$ inputs and Bob has one measurement ($Y=1$) with $B=X$ outcomes, the success rate $p_\text{mesd}=\frac{1}{X}\sum_{x}p(b=x|x)=\frac{D^2}{X}$ is achievable. In contrast, any strategy that involves separable measurements on $M$ and $B$ (including those that are adaptive) 
satisfies the bound $p_\text{mesd}\leq \frac{D}{X}$ (which is a generalization of Lemma~8 in \cite{Nathanson2005}). Indeed, writing $\tilde{\psi}_x^{MB}=\$_x\otimes \openone[\psi]$ and an arbitrary separable measurement as $B^{MB}_x=\sum_\lambda B_{x,\lambda}^M\otimes B_{x,\lambda}^B$, we find 
\begin{align}\nonumber
	p_{\text{mesd}}&=\frac{1}{X}\sum_{x,\lambda}\Tr\left(\tilde{\psi}_x^{MB}\left( B^{M}_{x,\lambda} \otimes B^B_{x,\lambda}\right)\right)\\
	&\leq \frac{1}{X}\sum_{x,\lambda}\Tr\left(\openone_M\otimes{\tilde\psi}^B\left( B^{M}_{x,\lambda} \otimes B^{B}_{x,\lambda}\right)\right)\nonumber\\
	&=\frac{1}{X}\Tr\left({\psi}^B \left(\openone^M\otimes \openone^B\right)\right) =\frac{D}{X},
\end{align}
where we used that $\tilde{\psi}^B=\Tr_M\left(\tilde{\psi}_x^{MB}\right)$ is independent of $x$. 

Even though joint measurements unleash the full power of shared entanglement, non-adaptive, product measurements can also be used to exploit entanglement and gain advantages over unassisted quantum communication. We give a striking example based on the RAC, with qubit communication. An unassisted qubit achieves at most $p^\text{qubit}_\text{win}=\frac{1}{2}\left(1+\frac{1}{\sqrt{2}}\right)$ \cite{Ambainis2008}. To beat this with a non-adaptive EA strategy, let the parties share $\ket{\phi^+}$ and let Alice apply the operation $U_{x_1x_2}=Z^{x_1}X^{x_2}$ on her local qubit before sending it to Bob. Bob is now in possession of one of the four Bell state $\ket{\phi_{x_1x_2}}=Z^{x_1}X^{x_2}\otimes \openone \ket{\phi^+}$. The first (second) bit is encoded in the phase (parity) of the Bell states. Thus, Bob can access the first (second) bit by measuring the phase (parity) observable $Z\otimes Z$ ($X\otimes X$) and achieve the perfect success rate $p_\text{win}=1$. This may be interpreted as a stochastic dense coding, in which both bits of Alice are in principle made available to Bob but only one of them is extracted (extracting both bits would require a joint phase and party measurement, which is non-separable). 
 
\textit{Discussion.} Our analysis of adaptive and non-adaptive protocols identifies different classes of quantum communication resources and begins to chart their landscape. The resource hierarchy may be illustrated as follows,
\[\begin{tikzcd}
	\text{EA qubit} \arrow[r] \arrow[rd] & \text{EA bit} \arrow[rd] \arrow[d,dashed,blue, bend right, "?"]  \arrow[r] & \text{qubit} \arrow[r] \arrow[d,dashed, bend right] & \text{bit}\\
	& \text{NA EA qubit} \arrow[r] \arrow[ru] \arrow[u,bend right,dashed] &\text{NA EA bit} \arrow[ru] \arrow[u,dashed, bend right]			
\end{tikzcd}\]
where a solid arrow indicates that one resource is strictly stronger than another, and a dashed one that there exist an example where one resource is stronger than the other.  The blue arrow with the question mark represents a relation yet to be settled: are there situations where an EA bit can outperform a non-adaptive EA qubit? We conjecture a positive answer. 

There are several interesting directions for future work. First, one could partition further (adaptive) EA qubit strategies in those that use joint measurements on the quantum message $M$ and the local system $B$, and those that use sequential measurement, where $M$ is first measured and the result used to select the measurements to be performed on $B$, or the other way around, and understand the relative power of such strategies.

Second, how does the above diagram have to be modified when considering non-binary messages, $D>2$? Although we found a strict resource relation between an EA bit and an unassisted qubit, the combined results of \cite{Tavakoli2021} and \cite{PawlowskiZukowski} show that there in general does not exist a strict ordering between these two resources for higher-dimensional messages.
 Nevertheless, as our result for $D=2$ illustrates, there may still be interesting classes of tasks for which strict resource relations do exist. 

Third, from the point of view of applications, is it possible to use adaptive EA classical protocols to boost the quantum reduction of communication complexity, in terms of the number of bits needed to deterministically compute a distributed function? Notably, most conventional approaches are based on non-adaptive EA classical protocols \cite{Review}. We have also seen that product measurements in EA quantum protocols are sufficient to outperform standard, unassisted, quantum communication. To what extent can this be leveraged for interesting quantum information protocols e.g.~in entanglement detection and cryptography? Finally, protocols with an adaptive advantage could be used to certify that a receiver maintains coherence up until receiving a message. This could be useful if the receiver is not trusted, and it would be interesting to investigate whether this relates to the security of cryptographic primitives like bit commitment.

\begin{acknowledgments}
We thank Anubhav Chaturvedi for pointing out the interpretation of adaptive protocols as Bell scenarios, Tam\'as V\'ertesi for sharing a proof of the projective bound for \eqref{facet} with us and letting us use this result, and  Mirjam Weilenmann and Miguel Navascu\'es for discussions. For the implementation of the NPA-type relaxations we acknowledge QETLAB \cite{qetlab}. This project was supported by the Wenner-Gren Foundations, by the EU Quantum Flagship Quantum Random Number Generators (QRANGE) project, by the Swiss National Science Foundation (Early Mobility Fellowship P2GEP2 188276),by the EU H2020 Quantum Flagship project QMiCS (820505), by the EU Quantum Flagship Quantum Random Number Generators (QRANGE) project, by the Fonds de la Recherche Scientifique – FNRS under Grant No PDR T.0171.22 and under grant Pint-Multi R.8014.21 as part of the QuantERA ERA-NET EU programme, by the FWO and the F.R.S.-FNRS under grant 40007526 of the Excellence of Science (EOS) programme. J. P. is a FRIA grantee and S.P. is a Senior Research Associate of the Fonds de la Recherche Scientifique - FNRS. 
\end{acknowledgments}

\bibliography{references_BellvsEACC_final}

\begin{thebibliography}{45}
\bibitem [{Not()}]{Note}%
  \BibitemOpen
  \bibinfo {title} {Specifically, one uses that \eqref{eq:bellcc} is invariant
  under the local relabelings $\{m\rightarrow \sigma(m) \,\, \& \,\,
  b_k\rightarrow b_{\sigma(k)}\}$, for any permutation $\sigma$ of the elements
  $[d]$, which allows one to reduce the number of independent variables in the
  moment matrix and the block-diagonalise it.}\BibitemShut {Stop}%
\end{thebibliography}
\end{document}